\documentclass[cits]{PoS}
\usepackage{graphicx} 
\usepackage{textgreek}
\usepackage{fontenc}
\usepackage{multicol}

\newcommand{\Ms}{M$_{\odot}$}
\newcommand{\env}{$\sim$}
\newcommand{\Ni}{$^{56}$Ni}
\newcommand{\Co}{$^{56}$Co}
\newcommand{\ens}{MgSiO$_3$}

\newcommand{\al}{Al$_2$O$_3$}

\title{Dust production in Supernovae}

\ShortTitle{Dust production in Supernovae}

\author{\speaker{Isabelle Cherchneff}\\
        Basel University\\
        E-mail: \email{isabelle.cherchneff@unibas.ch}}

\abstract{Supernovae have long been proposed to be efficient dust producers in galaxies. Observations in the mid-infrared indicate that dust forms a few hundred days after the stellar explosion. Yet, the chemical type and the amount of dust produced by supernovae are not well quantified. In this review, we summarise our current knowledge of dust formation derived from observations of supernovae, present the various theoretical models on dust synthesis and their predictions, and discuss these results in the context of the most recent observations of dust in supernova remnants.    }

\FullConference{The Life Cycle of Dust in the Universe: Observations, Theory, and Laboratory Experiments\\
		 18-22 November, 2013\\
		 Taipei, Taiwan}

\begin{document}

\section{Introduction}
\label{intro}
The ubiquity of cosmic dust in the local and far Universe poses the problem of identifying the sources of dust production. By cosmic dust, we mean solid grains of various chemical compositions, which include refractory compounds such as carbon (amorphous and graphite), metal oxides (alumina, fayalite, magnesia, spinel), silicates (amorphous or crystalline), silica, pure metals (iron, silicon or magnesium), metal sulphides, and other compounds that have still escaped identification \cite{mol10}. Dust in the dense Interstellar Medium is characterised by solid cores surrounded by icy mantles, which have grown in mass over time spans of the order of the molecular cloud lifetime. We know that dust grows in the cool part of the ISM but does not form from the gas phase, because the gas densities and temperatures characterising even the densest ISM regions are too low (typically, $10^4$ cm$^{-3}$~and $20$ K). The production of dust grains requires the high gas densities and temperatures encountered in the winds, outflows, or ejecta of evolved stars \cite {cher10a, cher13}. These sources of cosmic dust show an extreme variety of physical conditions, and include low-mass stars on the Asymptotic Giant Branch (AGB), massive stars in single or binary systems, such as supergiant stars, Large Blue Variables, and Wolf-Rayet stars, and rare stellar systems, such as the R Coronae Borealis stars. The explosion of evolved stars also produces dust grains. White Dwarves in binary systems and massive stars explode as Novae and Core-Collapse Supernovae (SNe), respectively, and dust has been detected in both environments. The common attributes to all these media are high gas densities and temperatures, and sufficient time to allow for the formation of molecules and large molecular clusters, i.e., dust clusters, from chemical processes. This stage corresponds to the dust nucleation phase. The time spent at high gas temperature and density should also permit the later coalescence and coagulation of dust clusters into solid grains, which corresponds to the dust condensation phase. Therefore, the synthesis of dust proceeds following these two steps for all dust forming environments in local or far-away galaxies. 

Not all stars contribute to the dust budget of a galaxy with similar importance. The paradigm that dominated thinking over the past decades was that AGB stars were the prevalent sources of galactic dust because they were numerous and efficient at forming solids in their deep inner winds \cite{gail99, cher12}. However, their contribution has been revised down following the recent surveys of AGB stars in local galaxies of low metallicity  conducted with the infrared (IR) space teslecope, Spitzer \cite{mat09, boy12}. Other potential dust providers to galaxies are the explosion of massive stars as core-collapse supernovae (SNe) \cite{hoy70}. More than two decades ago, dust production was observed a few hundred days after the explosion of the blue supergiant SK $-69^o202$ as SN1987A in the Large Magellanic Cloud \cite{luc89}. Dust has since been observed to form in several other SN ejecta \cite{ko05, ko06, ko09, gal12, sza13}, but the quantity of dust synthesised in these explosions is still debated. Should SNe be important dust producers, the solids grains produced in the ejecta evolve until the SN remnant (SNR) phase, and are processed by the reverse shock decades to centuries after their formation. So the final amount of ejecta dust that survives the SN remnant and is injected into the ISM is far from been settled. This value depends on the complex processes responsible for the formation of dust in SN ejecta, and the thermal and non-thermal sputtering of ejecta dust by shocks in SNRs. 

Large quantities of dust have been inferred to explain the reddening of quasar hosts at very high redshift. There exists no consensus on the ubiquity of dust in high-redshift galaxies, and several galaxies at $z> 6$ show no evidence of the cool dust that accompanies star formation in local galaxies \cite{fis14}. However, the large amount of dust inferred in the high redshift quasar J1148+5251 at $z=6.4$ and $t< 900$ million years after the Big Bang \cite{ber03}~forces us to identify potential candidates for dust factories in the early Universe. In this context, SNe may be key players to the dust budget in the primeval Universe, because of the short evolutionary time of their massive stellar progenitors. Furthermore, dust is a key ingredient to the cooling of primeval large gas structures and their subsequent mass fractionation, which leads to the formation of Population II stars at high redshift \cite{schnei12}. Hence, it is of paramount importance to understand the processes underpinning dust formation, and assess the amount of dust produced by SNe in local and far-away galaxies. 

In this review, we shall summarise the observations of dust in local SNe, review the existing theoretical efforts to model dust formation in SN ejecta, and present new theoretical developments on dust production. We finally discuss these results in the context of recent observations of dust in SNRs at far-IR and submillimetre (submm) wavelengths with Spitzer, AKARI, Herschel and ALMA. 
\section{Observational evidence of dust production in supernovae}
\label{obs}

\begin{figure} 
\includegraphics[width=.63\textwidth]{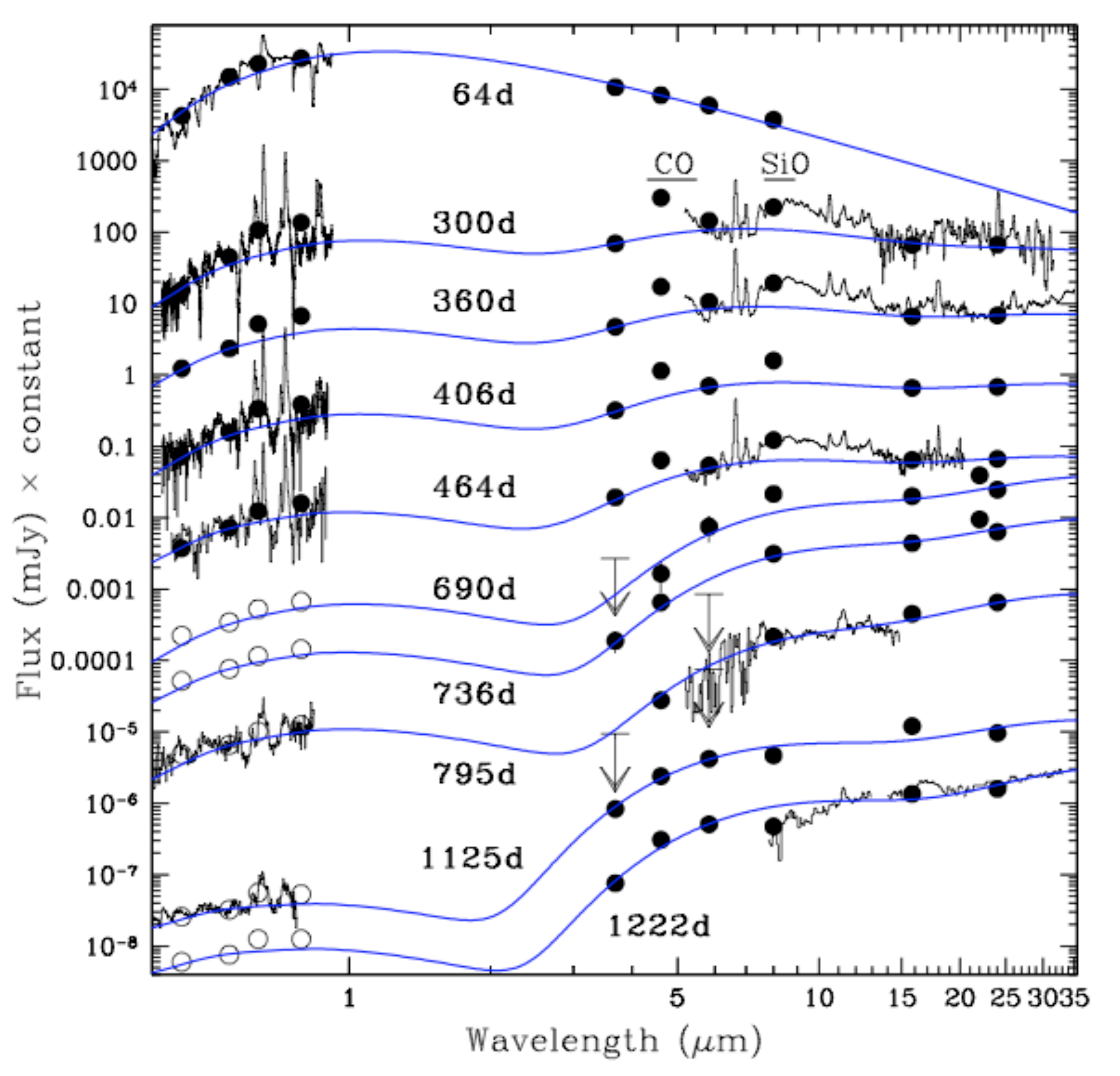}\centering
\caption{The mid-IR flux of the SN2004et ejecta at different epochs \cite{ko09}. The SiO and silicate dust bands are detected over the black-body continuum (blue solid line), and decrease in intensity with time.} 
\label{fig1}
\end{figure}

As mentioned in \S \ref{intro}, the best monitored dust formation event in a SN ejecta occurred in SN1987A in the Large Magellanic Cloud \cite{luc89}. The formation of dust in the ejecta is traced by 1) a sharp decline in the SN optical light curve at the onset of dust formation, 2) a blue-shift of the optical emission line profiles, likely due to attenuation by dust, and 3) an excess in the mid-IR spectral energy distribution of the SN, which implies that the dust detected has just formed from the ejecta gas phase, and is warm enough to emit at these wavelengths. These three imprints were observed in SN1987A around 500 days post-explosion. Vibrational transitions of molecules, that include the first overtone (2.3  $\mu$m) and fundamental (4.65  $\mu$m) bands of CO~\cite{spy88, cat88, dan91}, and the fundamental band of SiO at 8.1  $\mu$m \cite{roche91}, were detected \env 200 days post-explosion prior to the occurrence of dust. More recent observations of SNe confirm the presence of CO and SiO in the ejecta between 200 and 300 days post-outburst \cite{spy01, ko06, ko09, sza11, cher10b}. In SN2004et, the observed SiO mass gradually decreased over time, as illustrated in Figure \ref{fig1}, and could signal the onset of silicate dust synthesis in the ejecta \cite{ko09}. Recent ALMA data on SN1987A confirm the presence of large amounts of cool CO ascribed to the SN ejecta \cite{kam13}. 

The dust observed at mid-IR wavelengths a few hundred days after explosion forms in meagre amount, with masses that vary from \env $10^{-5}$ to $10^{-2}$~\Ms. These estimates are derived from fitting the IR excess and highly depend on radiative transfer modelling, the type of dust considered (usually amorphous carbon, AC, and interstellar silicates), and the degree of clumpiness of the ejecta. Indeed, a clumpy ejecta increases the derived dust mass by \env 50\% \cite{erc07}. However, a clear trend for rather modest masses of synthesised dust emerges from the modelling of mid-IR data at early post-explosion times. These results are in striking contrast with the dust masses derived in SNRs, decades to centuries after the initial explosion of the massive stellar progenitor. The SNR gas consists of dense ejecta clumps imbedded in a hot inter-clump medium. This evolved ejecta is crossed by a reverse shock, which develops when the mass of gas swept up by the forward explosion blast wave is as large or larger than the ejecta mass \cite{chev77}. Recent observations with Spitzer, Herschel, AKARI, and ALMA point to much larger masses of dust in SNRs \cite{gom14}. In Cas A, the remnant of a Type IIb SN, \env 0.025 \Ms\ of warm dust is detected at the reverse shock position with Spitzer \cite{rho08}. This warm dust traces material either in the post-reverse shock gas, or in the pre-shock gas, where dust is heated by the ultraviolet radiation present in the reverse shock precursor. A mass of tepid ejecta dust in the range of $0.06 - 0.085$~\Ms\ was also detected in the remnant with AKARI and Herschel \cite{sibt10, bar10}. In the Crab Nebula, which hosts a pulsar, \env $0.02-0.2$~\Ms\ of cool dust were deduced from the analysis of Spitzer and Herschel data \cite{gom12, tem13}. In SN1987A, a now young SNR, at least $0.2$~\Ms\ of ejecta dust was inferred from the analysis of Herschel and ALMA data \cite{mat11, inde14}. 

The ratio between SNR and SN dust masses thus ranges from $10$ to $10^4$. This discrepancy is puzzling and implies that either some cool dust was missed by the mid-IR observations, or dust continues to grow over time scales of decades after the explosion. The first scenario suggests that the dust formed at early times in the ejecta can cool very efficiently on very short time scale, perhaps if the ejecta is made of dense clumps. The second scenario is unlikely because the ejecta expends and cool as it evolves to the SNR phase. A diffuse and cool ejecta renders the nucleation and condensation of dust very difficult and inefficient. It was recently shown that no dust clusters could reform from the gas phase in SNRs \cite{bis14}. A third scenario, which is discussed in \S \ref{kin}, involves the gradual formation of dust in the nebular phase over a time span of a few years after explosion, when the gas is still warm and dense enough to allow for the efficient condensation of dust grains. 

\section{Types of dust}
\label{type}
The chemical type of the dust that forms in SN ejecta and the molecular precursors involved in the dust synthesis are not well characterised by observations, because high resolution spectroscopic data in the mid-IR are lacking. Three sub-categories of dust are usually taken into consideration, because they are part of the dust chemical families detected in other evolved circumstellar environments: 1) oxides; 2) carbon in the form of amorphous carbon and graphite; and 3) pure metal grains, where metal stands for iron, silicon, and magnesium in the case of SNe. The oxide family includes silicates of both pyroxene and olivine stoichiometry either in amorphous or crystalline form, silica, metal oxides, such as alumina and magnetite, and other oxides, such as magnesia. Several vibrational stretching and bending modes within the grain lattice give rise to emission or absorption bands in the mid-IR. These transitions then trace the chemical nature of their solid carrier. Dust species of relevance for circumstellar environments and their prevalent mid-IR spectroscopic signatures measured in the laboratory are summarised in Table 1. 

\begin{table}
\label{tab1}
\centering
\begin{tabular}{l r c l c}
\hline\hline
Family & Name & Formula & Bands & Reference\\
\hline
\multicolumn{5}{c}{Oxides} \\
\hline
Silicates&Pyroxene & Mg$_x$Fe$_{1-x}$SiO$_3$ & 10; 20  & \cite{dors95} \\
&Enstatite & {\ens} & $9-12$; 15.4; 19.5; 36.2& \cite{chi02}  \\
       & Ferrosilite & FeSiO$_3$ &11.3; 20.4; 31.7 & \cite{chi02} \\
& Olivine & Mg$_2x$Fe$_{2-2x}$SiO$_4$ &10; 20 & \cite{dors95} \\
 &Fosterite & Mg$_2$SiO$_4$ &  $10-12$; 16.3; $19.5- 24$ &\cite{koi03} \\
   &Fayalite & Fe$_2$SiO$_4$ & 10; $18-22$; 27; 32& \cite{suto02} \\
 Quartz &Silica & SiO$_2$ & 9.1; 12.6; $20.4-21.2 $; 26.1 & \cite{fab00} \\ 
Metal oxides & Alumina & Al$_2$O$_3$ &13 & \cite{koi95} \\
  &  Spinel & MgAl$_2$O$_4$ & 13; 16.8; 32 & \cite{fab01} \\
     & Magnesia & MgO & 19& \cite{hen95} \\
     & W{\"u}stite &FeO& 23.4 & \cite{hen95} \\
     & Hematite & Fe$_2$O$_3$ & 9.2; 18; 21; 20 & \cite{koi81} \\
     &Magnetite & Fe$_3$O$_4$ & 17; 25 & \cite{koi81} \\
     & Calcium oxide & CaO & 31.4 & \cite{hof03} \\
\hline
\multicolumn{5}{c}{Carbon} \\
\hline
&Amorphous carbon & C & 6.2; 8 & \cite{col95} \\
& Graphite  & C & 6.3; 11.52 & \cite{drai84}\\
\hline 
\multicolumn{5}{c}{Carbides} \\
\hline
 &Silicon carbide &  SiC& 11.3 &\cite{mut99} \\
& Titanium carbide & TiC& 20  & \cite{koid90} \\
\hline
\multicolumn{5}{c}{Sulphides} \\
\hline
 & Magnesium sulphide & MgS & 39 & \cite{hof03} \\
 & Iron sulphide & FeS & 34; 39  & \cite{hof03} \\
 \hline
\end{tabular}
\caption {Circumstellar dust types with typical band regions (in $\mu$m) in the mid-IR spectrum [$5-40$  $\mu$m] from laboratory data. }
\end{table}

\begin{figure} 
\includegraphics[width=.72\textwidth]{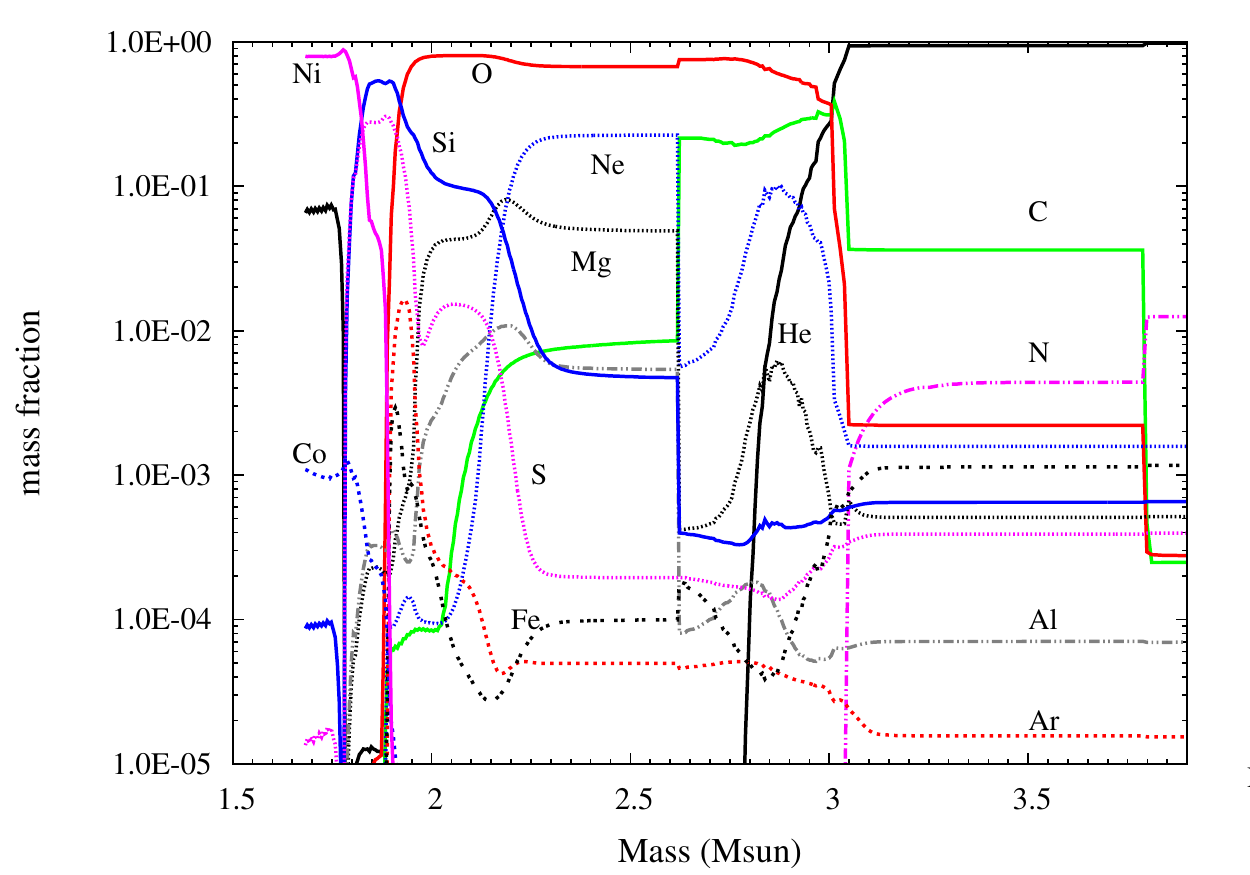}\centering
\caption{The helium-core zoning of a Type IIP SN with a 15 \Ms~stellar progenitor on the ZAMS. The yield data are taken from \cite{rau02}.} 
\label{fig2}
\end{figure}

It is still unclear what type of dust forms in SN ejecta. Existing theoretical models poorly constrain the chemical type of dust produced by SN ejecta. They often derive a dust composition on the basis of the chemical elements present in atomic form in the SN ejecta, and fail to provide a description of the chemical processes triggering dust synthesis. Recently, a new type of models have been developed, which involves a chemical kinetic approach to describe the nucleation phase of the dust production process. These models indicate alternative dust compositions, which are more limited than existing ones, because they reflect the control of the ejecta chemistry on the dust formation processes \cite{sar13}. 

A Type IIP SN ejecta is macroscopically mixed, owing to the  Raleigh-Taylor instabilities that develop a few hours after the SN explosion at the interface of the helium-core zones. The chemical composition of each zone depends on the nucleosynthesis experienced by the massive progenitor during its evolution, and the nucleosynthesis triggered by the SN explosion. As shown by several~2-~and~{3-D} explosion models, pockets of the hydrogen-rich, progenitor envelope are mixed inwards, while the products of explosion nucleosynthesis are expelled outwards \cite{ham10, jog13}. However, chemical mixing between zones on the microscopic scale does not occur. Thus, despite its disruption, the helium core retains some level of chemical stratification a few years after explosion. The elemental composition of the stratified SN ejecta stemming from the explosion of a 15 \Ms\ stellar progenitor is shown in Figure \ref{fig2}. The larger zone of the He-core is oxygen-rich, and includes large fractions of silicon, magnesium, and aluminium. Therefore, silica, magnesium-rich silicates, and alumina are expected to form in this zone. The O-rich zone is also characterised by a large mass of Ne and Ar atoms. The ionisation of these elements by the $\gamma$-ray photons induced by the decay of \Ni~and \Co, creates a population of inert gas ions, which impinges on the local chemistry of the gas phase. These ions efficiently destroy molecules, and thus delay the nucleation and condensation of dust grains \cite{sar13}. On the other hand, the outmost layers are rich in carbon and nitrogen, while the mass fraction of oxygen drops significantly. One should expect the production of carbon dust to prevail in this zone, accompanied with possible SiC grain formation. However, the zone is characterised by a large helium mass, and much like Ne and Ar, the ionisation of He in this zone delays the nucleation and condensation of dust grains to late post-explosion times ($>$ 1000 days) \cite{cher10c, sar13}. Compared with the 15 \Ms\ progenitor, more massive progenitors (15 \Ms\ < M$_{\star}$ < 30 \Ms) are characterised by a larger O-rich mass zone in the He-core, while less massive progenitors (8 \Ms\ < M$_{\star}$ < 15 \Ms) have a larger C-rich mass zone. 

To reproduce the observed mid-IR excess, radiative transfer models have highlighted various scenarios of dust compositions, essentially using mixtures of silicates and amorphous carbon grains. Depending on the object, this mixture is dominated by either silicates, e.g., SN2014et with its massive stellar progenitor \cite{ko09}, or carbon, like in SN2006BC, which has a small stellar progenitor (< 12 \Ms) \cite{gal12}. In the end, the chemical composition of the dust produced by SNe should primarily comprise silicates, silica, alumina, carbon, and silicon carbide, but iron sulphide and pure metal grains should also be present, albeit in smaller quantities. 

\section{Modelling dust formation in Supernovae}
\label{model}
As for other evolved circumstellar environments producing dust grains, the modelling of dust synthesis in SN ejecta usually involves a formalism based on Classical Nucleation Theory (or CNT), whereby the nucleation from the gas phase of a supersaturated compound occurs through the formation of critical clusters. This formalism was first developed to study the condensation of liquid water droplets under equilibrium in the Earth atmosphere \cite{fed66}. The CNT involves concepts like specific surface energy, supersaturation ratio, partial vapour pressure, and equilibrium between the supersaturated compound and its solid phase. However, the appropriateness of CNT has been questioned to describe dust synthesis in space \cite{don85}, where the formation of solids from the gas phase occurs in low-density, dynamical systems, which are out of equilibrium. The derived critical cluster size is usually on the atomic scale, and bulk property concepts, such as the surface surface energy, do not apply to small molecular clusters that are precursors to dust grains. Moreover, the existence of a steady-state critical cluster distribution is questionable when the molecular phase from which those clusters form is neither at equilibrium nor at steady state. Finally, the synthesis of ceramics and soot in flames in the laboratory have brought evidence of chemically and kinetically controlled processes out of equilibrium, which partake in the formation of solid particles. The synthesis of dust usually proceeds through a two-step mechanism, which entails the chemically-controlled nucleation of small dust clusters, and the growth of these clusters from surface deposition and coagulation during the condensation phase \cite{wool98}. The solids formed are characterised by amorphous or crystalline structures, depending on the gas physical parameters and the time spent by grains under these conditions.

\subsection{Existing models}
\label{old}

Dust production by SNe was first discussed by Clayton \cite{clay79}, while Kozasa et al. \cite{koz89} provided the first attempt to model the formation of dust in a SN ejecta, by studying SN1987A. The model was based on CNT applied to a homogeneous, fully-mixed ejecta wherein all chemical compounds were fully microscopically mixed. A similar formalism was used by Todini \& Ferrara \cite{tod01}, Nozawa et al. \cite{noz03}, Schneider et al. \cite{sch04} and Bianchi \& Schneider \cite{bian07} to describe the formation of dust grains in SN ejecta of various progenitor masses and metallicities, which include the massive Pair-Instability SNe present in the primeval Universe. These studies consider both fully-mixed and stratified SN ejecta, and assume the formation of dust out of the elemental gas composition. In these models, simple stoichiometric equations are used to describe the depletion of elements from the gas phase in solids of specific composition. For example, the synthesis of carbon is represented by the simple equation  $C_g = C_s$, while the production of silica is described by the equation $SiO_g + O_g = (SiO_2)_s$, where the subscripts $g$ and $s$ denote the gas and solid phase \cite{tod01, noz03, sch04, bian07}. However, these equations do not represent real chemical processes that partake in the formation of carbon and silica particles. It is well known that both the formation of carbon from a hydrogen-free environment and the production of silica through SiO vaporisation experiment involve complex chemical processes, which result in the formation of small fullerene cages and Si/O-rich small clusters, respectively \cite{dun12, reb08}. Some studies consider the formation of the key molecules CO and SiO by assuming their abundance is at steady state, but the non-equilibrium chemistry of the gas phase and the formation of other chemical species are not considered. The dust masses derived from all existing models are summarised in Table 2. Rather large masses of dust are obtained from these models as early as \env\ 400 days post-explosion, e.g., \cite{tod01}. 

These results are in contrast with the low dust mass values derived from IR observations, as discussed in \S\ \ref{obs}, because the CNT models fail to consider the nucleation phase. This phase is controlled by the local gas-phase chemistry, the formation of small dust clusters, the local physical processes at play (e.g., the $\gamma$-ray field), and the difference in physical conditions of the various dust-forming, ejecta mass zones. These combined mechanisms and conditions control and limit the occurrence of dust synthesis, and result in smaller dust mass yields. The nucleation phase is then a bottleneck to the condensation of solids in SN ejecta.   

\begin{table}
\centering
\begin{tabular}{lcccl}
\hline\hline
Model& Z&Fully Mixed& Progenitor& Total Dust \\
& & or Unmixed & Mass &  Mass \\
\hline
Kozasa et al. & Solar & FM & 19 \Ms&  -- \\
\cite{koz89}   & Solar  & U & 19 \Ms&  -- \\ 
    \hline
Kozasa et al.& Solar & U & 15 \Ms&  0.33 \Ms \\
\cite{koz09}    & Solar  & U & 20 \Ms&  0.68 \Ms \\ 

\hline
Todini \& Ferrara& Solar & FM & 12 \Ms& 0.20 \Ms\\
\cite{tod01} & Solar  & FM & 15 \Ms  &0.45 \Ms\\
& Solar & FM &  20 \Ms &0.70 \Ms\\ 
& Solar  & FM &  25 \Ms   &1.00 \Ms\\ 
& 0 & FM &  15 \Ms& 0.45 \Ms\\ 
& 0 & FM &  20 \Ms&0.08 \Ms\\ 
& 0 & FM &  25 \Ms &0.08 \Ms\\ 
\hline
Nozawa et al.  &  0& FM & 20 \Ms & 0.73 \Ms \\
 \cite{noz03}&  0& U & 20 \Ms & 0.57 \Ms \\
 \hline
Bianchi \& Schneider &  Solar& FM & 12 \Ms & 0.12 \Ms \\
 \cite{bian07} &  Solar & FM & 15 \Ms &0.28 \Ms \\
 & Solar& FM  & 20 \Ms & 0.40 \Ms \\
 &  Solar& FM  & 25 \Ms & 0.62 \Ms \\
\hline
Cherchneff \& Dwek~$^a$& 0& FM & 20 \Ms & 0.16 \Ms \\
\cite{cher10c} & 0 & U & 20 \Ms &  0.10 \Ms \\
\hline
Sarangi \& Cherchneff~$^a$& Solar & U & 12 \Ms  & 0.048 \Ms \\
 \cite{sar13}& Solar & U & 15 \Ms  & 0.038 \Ms \\
  & Solar & U & 19 \Ms & 0.035 \Ms \\
  & Solar & U & 25 \Ms & 0.09 \Ms \\
  \hline
\end{tabular}
\caption{Dust mass estimated by existing dust formation models for metallicity Z = Z$_{solar}$ and Z = 0 (adapted from \cite{sar13}.$^a$ Chemical kinetic model.}
\label{tab2}
\end{table}

\subsection{Chemical kinetics models}
\label{kin}

\begin{figure} 
\includegraphics[width=.65\textwidth]{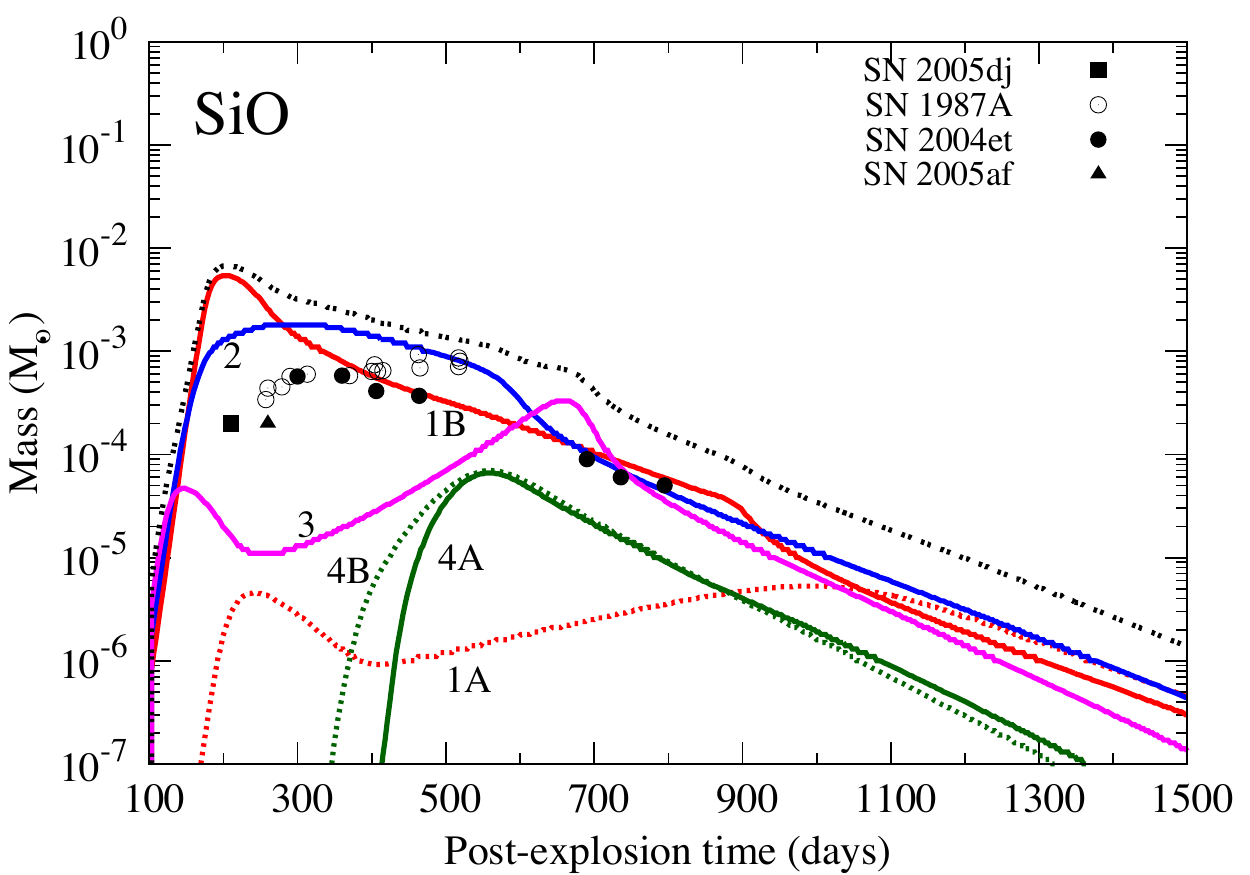}\centering\\
\includegraphics[width=.65\textwidth]{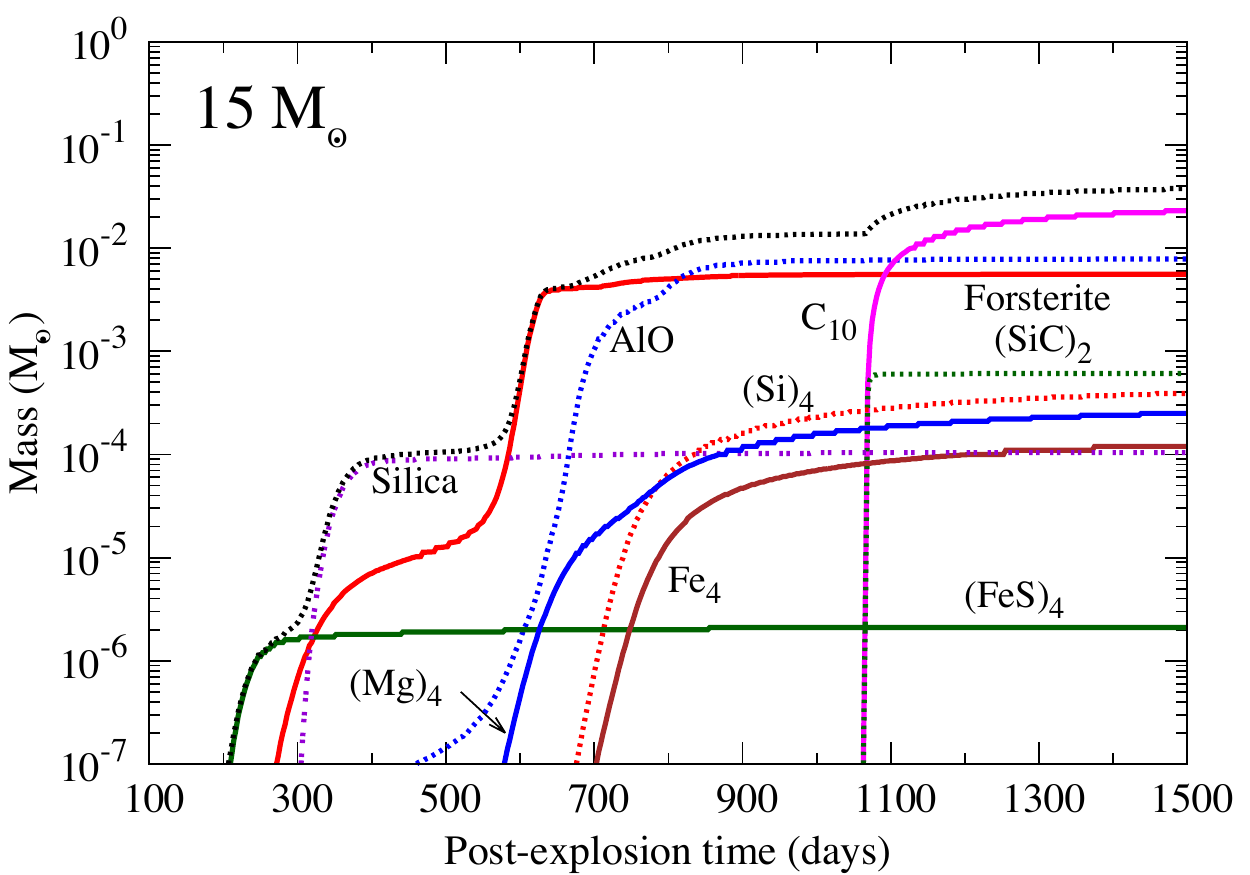}\centering
\caption{The SiO mass for the various ejecta mass zones (top) and the dust cluster mass summed over all ejecta zones (bottom) as a function of post-explosion time for a SN with a 15 \Ms\ stellar progenitor \cite{sar13}. Symbols in the top figure are for SiO masses derived from IR data for several SNe.} 
\label{fig3}
\end{figure}

A new approach, which treats both the nucleation and the condensation phases that characterise dust production in evolved circumstellar media, has recently been developed and applied to the ejecta of primeval, massive SN explosions and local core-collapse SNe \cite{cher09, cher10c, sar13}. Unlike other models, these new studies include the formation of molecules and dust clusters by all relevant chemical reactions in the gas phase, thereby limiting the number of elements that can be depleted in dust grains. The chemical description makes use of all thermal processes, such as termolecular, bimolecular, and unimolecular reactions and their reverse processes, and non-thermal processes, i.e., the ionisation by Compton electrons produced from the decay of radioactivity-induced $\gamma$ rays in the ejecta. The synthesis of dust clusters occurs following a series of chemical processes that build up the first cluster structures, which are able to later condense into larger grains. In the case of silicates, the formation of enstatite and forsterite dimers is described by the slow dimerisation of SiO as a first step. It then proceeds by a series of recurrent reactions, which includes the oxygen addition step through the reaction with O-bearing species like O$_2$ or SO, and the inclusion of one Mg atom in the cluster lattice. This nucleation scheme was proposed by Goumans \& Bromley \cite{gou12} to model the formation of silicates in AGB stellar winds. In the case of carbon, the lack of microscopically-mixed hydrogen in the carbon-rich, outer zone of the ejecta hampers the formation of amorphous carbon through the synthesis of polycyclic aromatics hydrocarbons and grapheme sheets \cite{cher11}. Instead, the growth of carbon proceeds through the formation of pure carbon chains and rings, until large monocyclic rings close to form the first stable carbon cage C$_{28}$ \cite{dun12}. The growth of alumina is described following the dimerisation of the molecule AlO, and the oxidation of AlO dimers to \al\ through reactions with O-bearing species \cite{bis14}. Finally, for pure metal clusters, the nucleation phase corresponds to a series of chemical reactions that form small M$_4$ tetramers (where M is the metal).  

The challenge for these models is to reproduce both the observational data on molecules, such as CO and SiO, and the dust mass produced in the nebular phase. Results on SiO and dust cluster masses formed in the ejecta of a SN with a 15 \Ms\ stellar progenitor are shown in Fig \ref{fig3} \cite{sar13}. Silicon monoxide forms early on in the various O-rich zones of the ejecta (labelled 1B, 2, 3, 4A, and 4B), but it is quickly depleted in SiO dimers, silica, and small silicate clusters after 300 days. The decline in SiO mass with time is observed in a few SNe, and the chemical kinetic model reproduces this trend. Another key molecule is CO. The molecule primarily forms in the O-rich zones 4A and 4B, where the CO mass increases from \env $10^{-3}$ \Ms\ at day 150 to $0.1$ \Ms\ at day 1500. Interestingly, zones 4A and 4B form very small amounts of dust clusters. Therefore, CO is not a dust tracer, but a resilient, free-flying molecule that evolves to the SN remnant phase with large masses. Recent observations of SN1987A with ALMA indeed indicate large masses  ($> 10^{-2}$ \Ms) of cool, ejecta CO at the centre of the young remnant \cite{kam13}. 

As for dust clusters, their growth is gradual, and follows a series of cluster formation events, which occur in different ejecta zones. At the times covered by mid-IR observations, the dust cluster mass is rather modest and ranges between $10^{-6}$ and $10^{-3}$. This value range agrees with the dust masses derived from observations. At later times, the formation of alumina, silicates of forsterite stoichiometry, and finally carbon takes place, and increases the cluster mass up to \env\ $4\times 10^{-2}$ \Ms. The total cluster masses derived by chemical kinetic models for several stellar progenitors are listed in Table 2. These values are smaller than the dust mass derived from CNT models by a factor of \env\ 10, and they are close to the dust masses inferred from the latest submm observations of SN remnants (see \S \ref{insight}). Then, this gradual trend in dust growth over a time span of a few years after the SN explosion represents a genuine explanation to the discrepancy on dust mass derived from IR and submm data of SN and SN remnants. 

\begin{figure} 
\includegraphics[width=.73\textwidth]{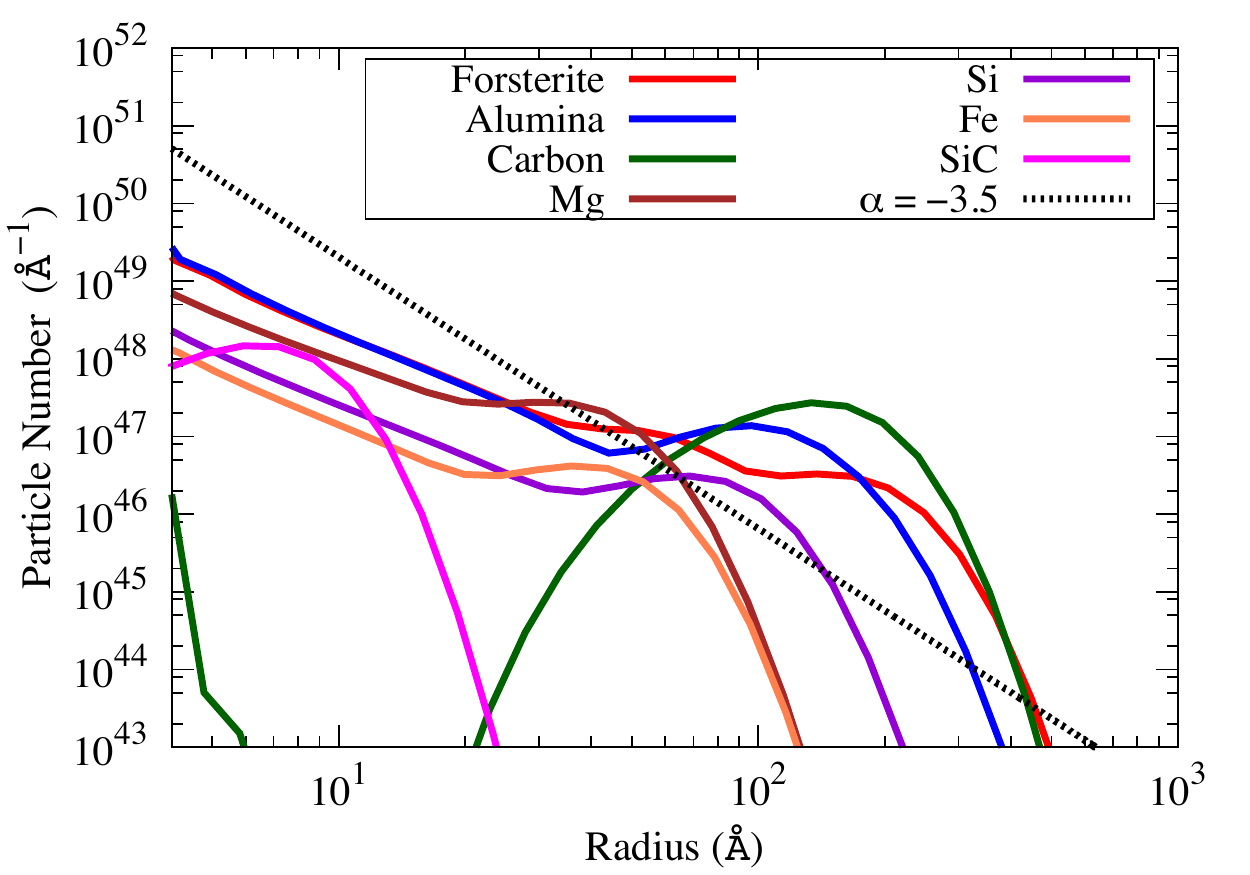}\centering
\caption{Grain size distributions for various types of dust formed at 2000 days post-explosion in the SN ejecta associated with a 15 \Ms\ stellar progenitor. The corresponding MRN distribution with exponent -3.5 for the total number of grains formed in the ejecta is shown for comparison \cite{sar14}. } 
\label{fig4}
\end{figure}

The most recent chemical kinetic models on dust formation in Type II-P SN ejecta now couple the chemically-controlled nucleation phase to the dust condensation phase. The condensation phase involves the coagulation of small clusters owing to Brownian diffusion, which accounts for the scattering, collision and coalescence of the grains through Brownian motion \cite{sar14}. They provide grain size distributions for the various types of solids produced. The dust size distributions for the homogeneous, stratified ejecta of a SN associated with a 15 \Ms\ stellar progenitor is shown in Fig \ref{fig4}. The size distributions are dominated by very small grains with radii smaller than $10$~{\AA}, but there is also a significant number of large grains with radii that range from 100 to 1000 {\AA}, for carbon, silicate, and alumina dust. The pure metal grains have smaller size distributions peaking at \env\ $40-80$ {\AA}. The grain size distributions are in all cases different from a MRN distribution usually assumed for the description of dust in the Interstellar Medium \cite{mat77}. The total dust mass formed in this model is $0.035$ \Ms\ \cite{sar13}. When a clumpy ejecta is considered, larger grains are formed and reach the $1 \mu$m size. The total dust mass is also larger by a factor of \env\ 3 and reaches \env\ 0.1 \Ms\ \cite{sar14}. 

A recent study of the SN progenitor that led to the SNR Cas A highlights the dependence on gas density of the dust chemical nature \cite{bis14}. Cas A is the remnant of a Type IIb SN explosion, whose ejecta is characterised by a low-density gas, as the stellar progenitor had lost most of its hydrogen envelope prior to explosion. A chemical kinetic model results in almost no dust formed in the ejecta, because the gas is too diffuse. This result is in contrast with existing dust condensation models based on CNT \cite{noz10}, and Spitzer observations of Cas A, whose analysis requires several grain populations of various chemical types \cite{rho08, ar14}. The dust mass as a function of increase in gas number density in the SN ejecta that led to Cas~A is illustrated in Figure \ref{fig4} for various chemical types of solids \cite{bis14}. The type of dust that forms in SN ejecta reflects the physical conditions of their birth site, and the larger the gas density, the greater the chemical complexity of the dust that forms. Carbon dust is particularly dependent on gas density, and forms in the densest ejecta, that are typical of Type II-P SNe. Therefore, chemical kinetic models indirectly point to a clumpy ejecta for the Cas~A progenitor, to reproduce mid-IR data. These clumps are indeed observed in the Cas~A remnant as extended filaments \cite{fes06}.  

\begin{figure} 
\includegraphics[width=.74\textwidth]{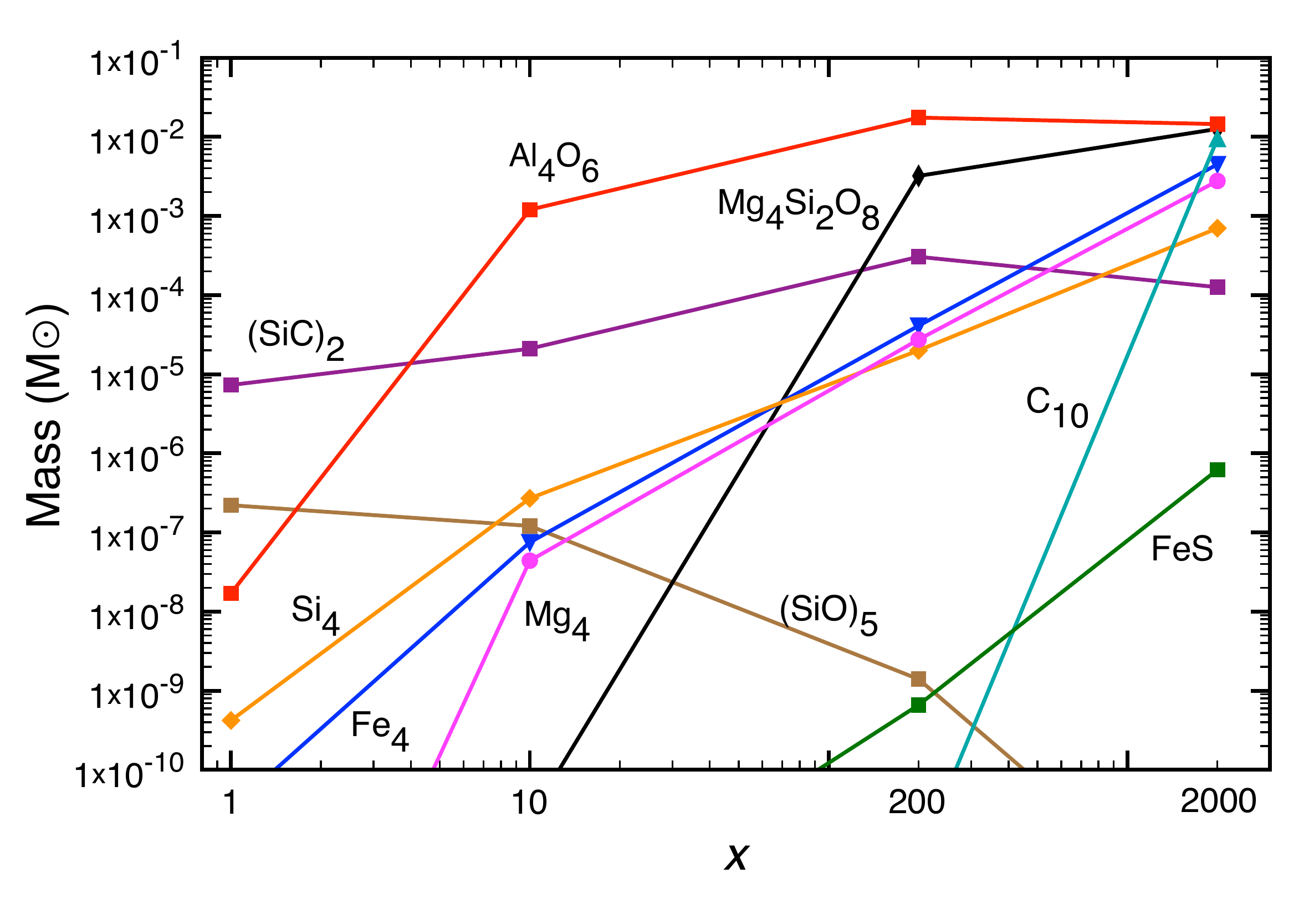}\centering
\caption{Dust cluster types for a  SN with 19 \Ms\ progenitor, as a function of ejecta density \cite{bis14}. } 
\label{fig5}
\end{figure}

\section{New insights from supernova remnants}
\label{insight}

Recent observations of dust in SN remnants with Spitzer, Herschel, and ALMA have rekindled the debate on dust synthesis in SNe, and whether these objects are major contributors to the total dust budget of galaxies. While Spitzer observations probe the warm dust content of SN remnants probably related to the reverse shock crossing the ejecta, the far-IR and submm data obtained with Herschel and ALMA trace the cool dust that possibly formed in the SN ejecta. The later observations indicate large masses of cool dust in several SNRs. The most studied SN remnants and their estimated dust mass are listed in Table 3. 
%
\begin{table}
\centering
\begin{tabular}{lcccc}
\hline\hline
 Name & SN1987A& Cas A & \multicolumn{2}{c}{Crab Nebula} \\
Age (yr) & 27 & \env\ 330 & \multicolumn{2}{c}{\env\ 960} \\
Progenitor mass (\Ms)  & 19 & 19 & \multicolumn{2}{c}{12} \\
Data & Herschel $-$ ALMA& Spitzer $-$ Herschel & \multicolumn{2}{c}{Spitzer $-$ Herschel}\\
Dust type & Carbon, silicate, iron & Silicate &\multicolumn{2}{c}{Silicate \& carbon}  \\
Dust Mass (\Ms)& $\ge 0.2$ &  \env\ 0.1  & 0.24 - 0.11 & 0.13 - 0.03 \\
Dust Temp. (K) & 17 - 23 & 22 - 26 & 35  &  35 - 68 \\
Reference & \cite{mat11,inde14}&  \cite{rho08,bar10} &  \cite{gom12} &\cite{tem13}  \\
\hline
\end{tabular} 
\caption{Dust chemical composition, temperature range and mass derived for various SN remnants.}
\end{table} 
 
The data are reproduced by assuming a simple chemical composition of dust grains, usually amorphous carbon or a combination of amorphous carbon and silicate, and a single emission temperature depending on the type of dust grains, e.g., \cite{mat11, bar10, gom12, inde14}. These models omit the various dust chemical families that should form in SN ejecta (e.g., alumina or pure metal grains). Furthermore, the ejecta being clumpy, each clump forms dust according to its specific chemical composition and physical conditions. However, all clumps forming a specific type of dust, such as silicate, alumina or amorphous carbon, are not equal in terms of initial chemical composition, $\gamma$-ray field, gas temperature, and density. In the case of the Crab Nebula, Temim \& Dwek \cite{tem13} model Spitzer and Herschel data and study the impact of varying the dust size distributions, which are characterised by different dust temperatures combined with using grain absorption coefficients with a more complete wavelength coverage. The resulting dust mass is reduced by a factor of 2 to 4 compared to a single-temperature dust model - see Table 3. 

Therefore, the dust mass inferred from IR and submm data of SNRs is likely to be revised, as more ALMA data, chemical kinetic studies of dust synthesis, and sophisticated radiative transfer models are made available.

\section{Conclusions}
\label{conc}

The dust production by SNe still raises many questions. The current paradigm on the mass of dust formed in the nebular phase converges towards fair amounts of dust synthesised, with typical dust mass yields in the range $10^{-2}-2\times 10^{-1}$ \Ms. These values are smaller than the \env\ 1 \Ms\ value required to explain the large quantities of dust in young galaxies \cite{dwek07}. However, primitive SNe originate from stellar progenitors more massive than 25 \Ms\ and may thus produce dust yields at least as large as 0.2 \Ms. 

The evolution of the dust in the remnant remains unclear. Several studies tackle the sputtering by the reverse shock of ejecta grains in SNRs. The models either consider homogeneous SNR gas and thermal sputtering \cite{noz07, noz10}, or ejecta clumps that are disrupted over several clump-crushing times, and the subsequent thermal sputtering in the hot inter-clump gas during disruption \cite{sil10}. Both models consider ejecta grain size distributions derived from CNT by Nozawa et al.  \cite{noz03} for primeval core-collapse SNe, which could be well fitted by a combination of power-law functions of exponent $-3.5$ and $-2.5$. As we previously mentioned in \S \ref{kin}, the size distributions of grains produced in the SN ejecta may greatly differ from a MRN distribution typical of interstellar dust, and distributions derived from CNT. Both models derive dust mass survival rates in the range $10-80$ \%, and these rates depend on the dust chemical composition and initial size distribution, with a higher survival rate for larger grains. 

Recent modelling efforts consider the chemistry of a clump crossed by the reverse shock in Cas A, and show that, once destroyed, the dust is unable to reform out of the gas phase in the immediate post-shock gas and over time spans of the order of the clump-crushing time \cite{bis14}. It appears that any small dust sputtered by the reverse shock in SNRs will then be lost to the total SN dust budget. The survival of large dust grains will depend on their chemical composition, size, and location in the SNR. The disruption of dense clumps induced by the crossing of the reverse shock in the remnant results in high density regions or bullets travelling in the SNR \cite{sil10}. These inhomogeneities could be the ideal place where large, SN ejecta grains would escape harsh thermal sputtering, and be injected into the Interstellar Medium. This scenario is supported by the isotopic analysis of pre-solar SN grains in meteorites \cite{zin07}, which highlight the presence of large supernova dust grains that formed a few years after explosion and eventually make their way to our Solar System and to Earth.

\end{document}